\documentclass[american,english,aps,twocolumn,groupaddress,showpacs,prl]{revtex4}
\usepackage[T1]{fontenc}
\usepackage[latin9]{inputenc}
\usepackage{amstext}
\usepackage{graphicx}

\makeatletter
\@ifundefined{textcolor}{}
{%
 \definecolor{BLACK}{gray}{0}
 \definecolor{WHITE}{gray}{1}
 \definecolor{RED}{rgb}{1,0,0}
 \definecolor{GREEN}{rgb}{0,1,0}
 \definecolor{BLUE}{rgb}{0,0,1}
 \definecolor{CYAN}{cmyk}{1,0,0,0}
 \definecolor{MAGENTA}{cmyk}{0,1,0,0}
 \definecolor{YELLOW}{cmyk}{0,0,1,0}
 }

\usepackage{times}
\renewcommand{\citet}{\cite}
\usepackage{mathptmx}
\usepackage{graphicx}
\usepackage{bm}

\makeatother

\usepackage{babel}

\begin{document}

\title{Thermal rectification in asymmetric U-shaped graphene flakes}

\author{Jigger Cheh$^{1}$}

\email{jk_jigger@xmu.edu.cn}

\author{Hong Zhao$^{1,2}$}

\email{zhaoh@xmu.edu.cn}

\affiliation{$^{1}$Department of Physics, Institute of Theoretical Physics and
Astrophysics, Xiamen University, Xiamen 361005, China}

\affiliation{$^{2}$State Key Laboratory for Nonlinear Mechanics, Institute of
Mechanics, Chinese Academy of Sciences, Beijing 100080, China}

\begin{abstract}
In this paper, we study the thermal rectification in asymmetric U-shaped
graphene flakes by using nonequilibrium molecular dynamics simulations.
The graphene flakes are composed by a beam and two arms. It is found
that the heat flux runs preferentially from the wide arm to the narrow
arm which indicates a strong rectification effect. The dependence
of the rectification ratio upon the heat flux, the length and the
width of the beam, the length and width of the two arms are studied.
The result suggests a possible route to manage heat dissipation in
U-shaped graphene based nanoelectronic devices.
\end{abstract}
\maketitle
Graphene, a single layer of carbon atoms arranged in a honeycomb lattice,
has attracted much interest due to its extraordinary properties\citet{01.graphene1,02.graphene2}.
Since graphene exhibits much greater electron mobility than silicon
as a zero band gap semiconductor, it has been considered as a promising
candidate for the post-CMOS (complementary metal-oxide-semiconductor)
material to replace silicon which is approaching its fundamental limit\citet{03.cmos}.
As electronic devices would undergo dramatic miniaturization, thus
heat dissipation has become one of the most important barriers of
breaking through. To achieve better functionality and longer lifetime
for nanoelectronic devices, it is desirable to have in-depth understanding
of the thermal properties of graphene which stimulates intense efforts
both experimentally\citet{04.exp1,05.exp2,06.exp3} and theoretically\citet{07.t1,08.t2}.
To design a nanoelectronic device with better heat dissipation capacity,
one of the most challenging issues is to design thermal rectifiers.
Thermal rectification is a phenomenon that the heat flux runs preferentially
in one direction and inferiorly in the opposite direction\citet{09.rect00.cw.chang,10.rect01}.
Thus realization thermal rectification in graphene has deep implication
for graphene based devices. Through molecular dynamics simulations,
researchers have proposed several different thermal rectifiers from
the asymmetric graphene nanoribbons\citet{11.bb1,12.bb2,13.bb3,14.bb4}.
In nanoelectronic designs, U-shaped devices are very common and widely
used as electronic transistors and logic gates. Very recently it is
found that the U-shaped graphene flakes reveal extremely high I{\footnotesize on}/I{\footnotesize off}
ratio as channel transistors in experiments and they can easily realize
and control the resonant tunneling without any external gates\citet{15.U-1,16.U-2}.
The U-shaped graphene flakes can be fabricated by using lithography
or gallium focused ion beam to cut from continuous graphene sheets\citet{15.U-1,16.U-2}.
Therefore it arouses great interest to design thermal rectifiers by
U-shaped graphene flakes.

Here we study the thermal rectification in asymmetric U-shaped graphene
flakes by NEMD (nonequilibrium molecular dynamics) simulations. The
graphene flakes are composed by a beam and two arms. We report that
higher thermal conductivity is obtained when the heat flux runs from
the wide arm to the narrow arm. We also discuss the impacts of the
heat flux, the length and the width of the beam, the length and width
of the two arms on the rectification ratio. Our result may inspire
the experimentalists to realize thermal rectification in U-shaped
graphene flakes.%
\begin{figure}
\includegraphics[scale=0.18]{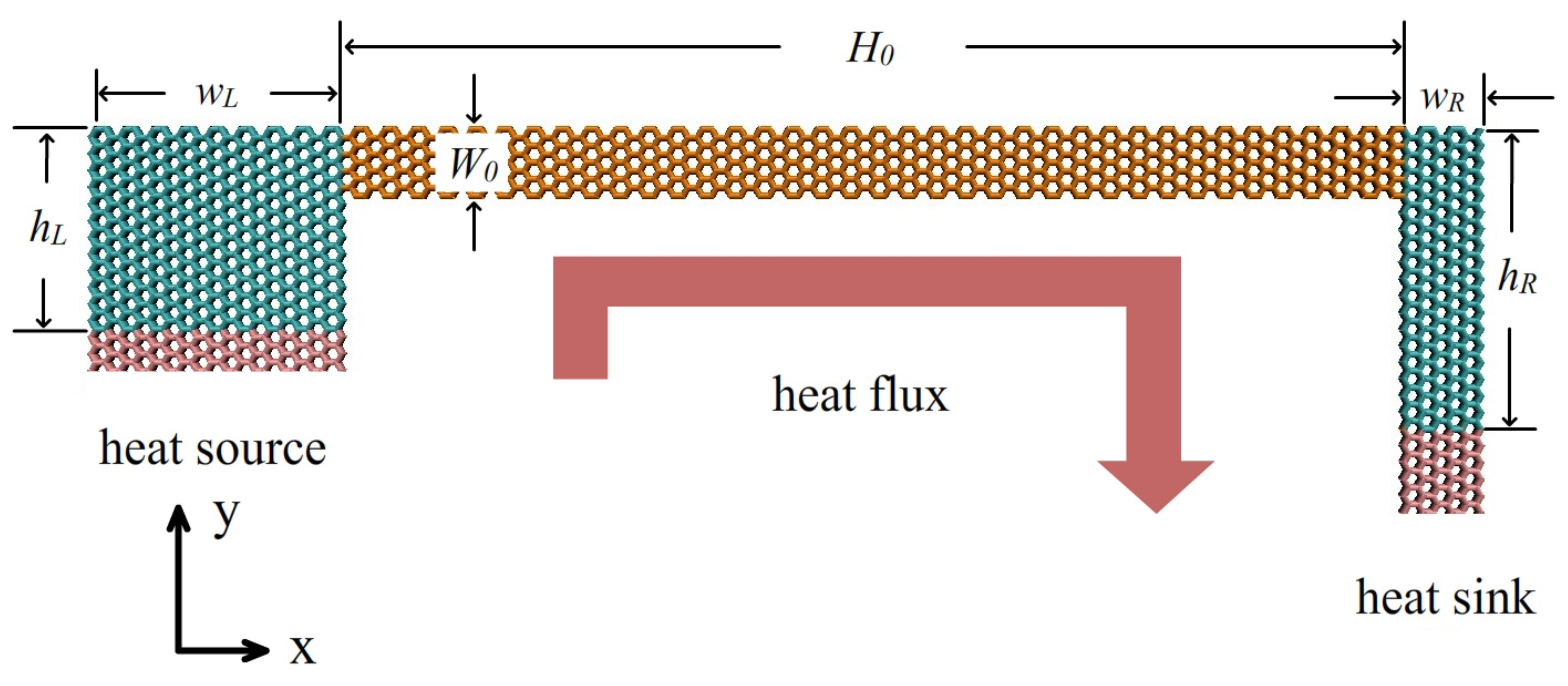}

\caption{(Color online) Schematic of the U-shaped graphene flakes. The graphene
flakes are composed by a beam and two arms. The heat source and heat
sink are connected to the two arms. The heat flux runs from the heat
source to the heat sink.}

\end{figure}

In Fig. 1 we show the structure of the U-shaped graphene flakes. The
graphene flakes are composed by a beam and two arms. The beam is drawn
in orange with a length of $H_{0}$ and a width of $W_{0}$. Different
$H_{0}$ and $W_{0}$ are applied to investigate the dependence of
thermal rectification ratio upon the size of the beam. The left arm
is drawn in cyan with a length of $h_{L}$ and a width of $w_{L}$.
The right arm is also drawn in cyan with a length of $h_{R}$ and
a width of $w_{R}$. Different ratio of $h_{R}/h_{L}$ and $w_{R}/w_{L}$
are applied to investigate the dependence of thermal rectification
upon the structural asymmetry. The end of the two arms is connected
with either the heat source or the heat sink. The heat source and
heat sink are drawn in red. Their outmost edges are frozen which is
corresponding to fixed boundary condition. The heat flux runs from
the heat source to the heat sink. We use the same reactive empirical
bond-order (REBO) potential\citet{17.rebo} as implemented in the
LAMMPS\citet{18.lammps} code to simulate the anharmonic coupling
between the carbon atoms. Equations of motions are integrated with
velocity Verlet algorithm with the minimum timestep $\bigtriangleup t=0.25$
fs.

First the graphene flakes are equilibrated at a constant temperature
$T=300$ K in the Nose-Hoover thermostat by 0.75 ns. After that the
heat flux is imposed. It is realized by the energy and momentum conserving
velocity rescaling algorithm developed by Jude and Jullien\citet{19.jj}.
It is widely used to investigate thermal rectification in different
materials\citet{20.jj-r1,21.jj-r2}. By rescaling atomic velocities
at each time step $dt$, specific amount of kinetic energy $dE$ is
added in the heat source and removed in the heat sink respectively.
The heat flux can be calculated by $J=dE/dt$. The temperature profiles
of the beam and the two arms are obtained by dividing the graphene
flakes by several slabs of a constant length 4 \foreignlanguage{american}{$\textrm{\AA}$}
along their axis respectively. The local temperature of each slab
is derived from the averaged kinetic energy. We average the temperature
profiles over 100 ps after the heat flux is imposed. After 2 ns the
temperature profiles do not change much and the whole nonequilibrium
simulation process covers 3 ns.%
\begin{figure}
\includegraphics[scale=0.24]{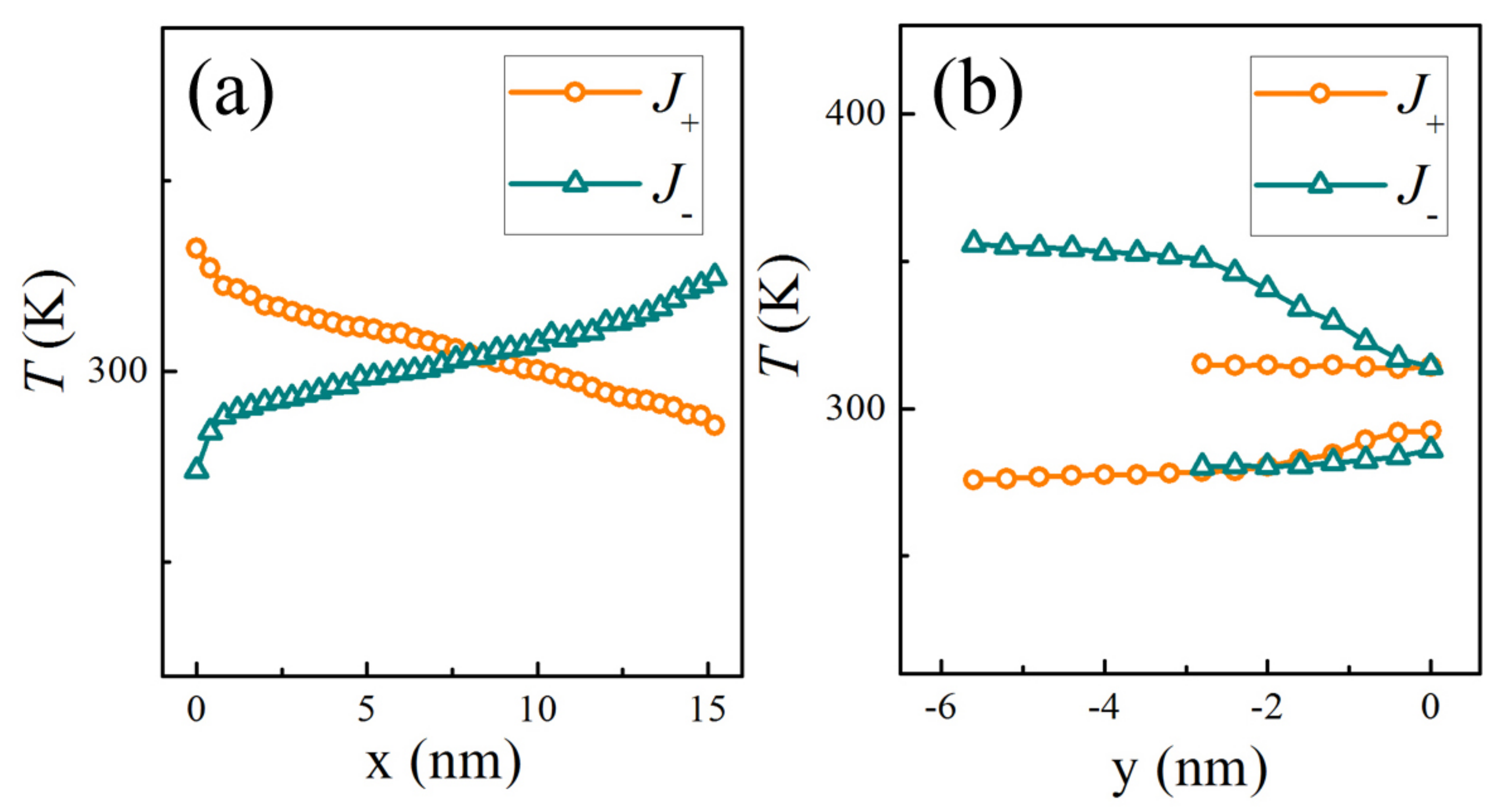}\caption{(Color online) (a) Typical temperature profiles on the beam. (b) Typical
temperature profiles on the two arms. The left arm is shorter than
the right arm.}

\end{figure}
\begin{figure}
\includegraphics[scale=0.24]{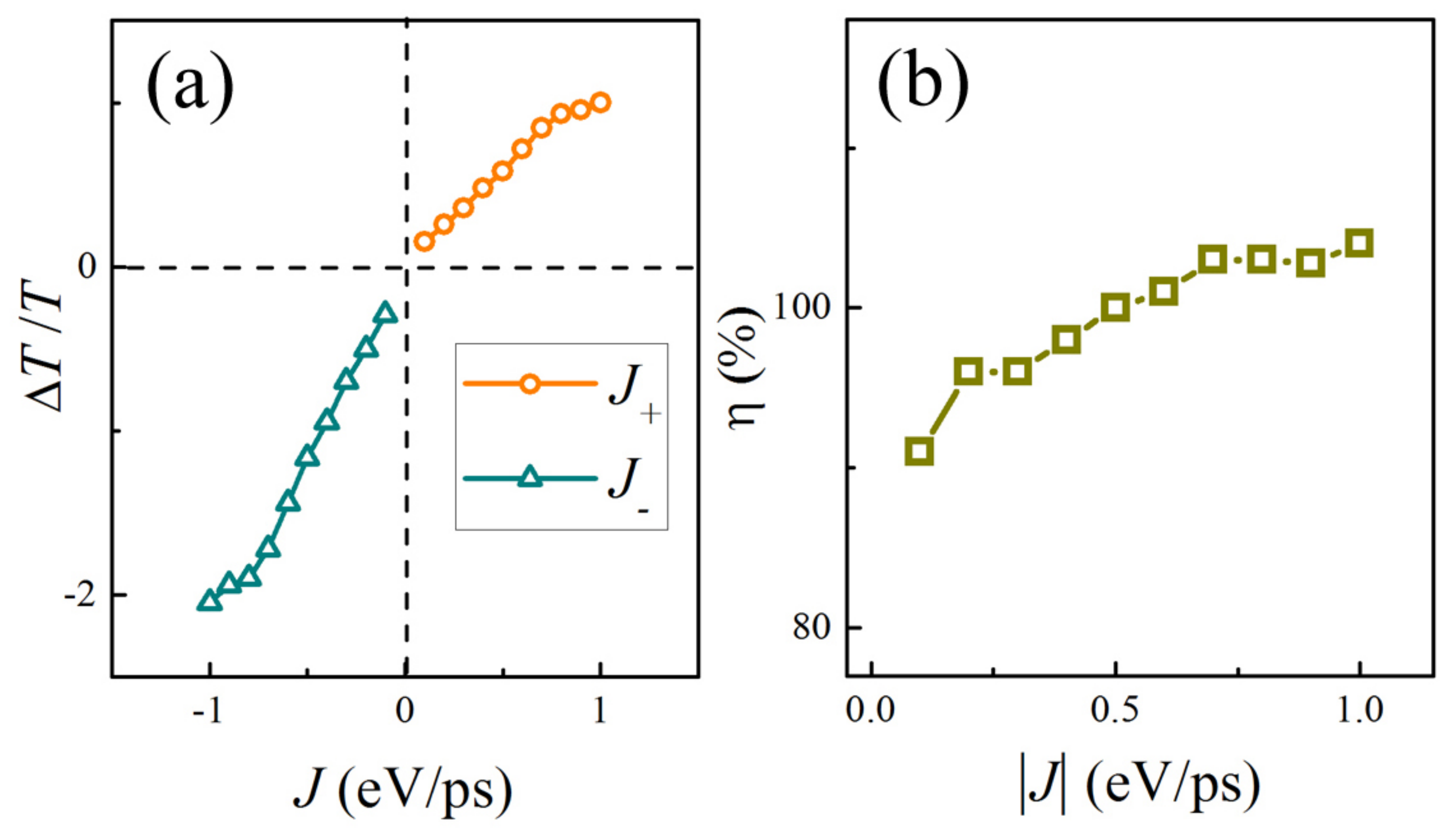}

\caption{(Color online) (a) Heat flux $J$ vs temperature differences $\bigtriangleup T/T$
between the two arms. $J_{+}$ corresponds to $\bigtriangleup T_{+}>0$
and $J_{-}$ corresponds to $\bigtriangleup T_{-}<0$. (b) Thermal
rectification ratio $\eta$ vs heat flux $J$.}

\end{figure}

The heat flux runs from the heat source to the heat sink along the
U-shaped graphene flakes. We label the heat flux as $J_{+}$ ($J_{+}>0$)
when the heat source is connected to the left arm and the heat sink
is connected to the right arm. Similarly we label the reversed heat
flux as $J_{-}$ ($J_{-}<0$) when the heat source is connected to
the right arm and the heat sink is connected to the left arm. Due
to the thermal rectification effect, different temperature profiles
would be obtained by reversing the heat flux. According the Fourier's
Law, thermal conductivity of the U-shaped graphene flakes can be qualified
as: 

\begin{equation}
G_{+}=\frac{J_{+}/A}{\bigtriangleup T_{+}/L}\qquad G_{-}=\frac{J_{-}/A}{\bigtriangleup T_{-}/L}\end{equation}

\noindent Here $A$ is the averaged cross section, $L$ is the distance
between the two arms, $\bigtriangleup T_{+}$ ($\bigtriangleup T_{-}$)
is the temperature difference between the left arm and the right arm
when $J_{+}$ ($J_{-}$) is imposed. Since $J_{+}=-J_{-}$, $A$ and
$L$ are the same, so the thermal rectification ratio $\eta$ is qualified
as\citet{09.rect00.cw.chang,20.jj-r1,21.jj-r2,22.mp2}:

\noindent \begin{equation}
\eta=\frac{G_{+}-G_{-}}{G_{-}}=(\frac{-\bigtriangleup T_{-}}{\bigtriangleup T_{+}}-1)\times100\%\end{equation}

In Fig. 2 we show the typical temperature profiles on the beam and
the two arms. Here $J_{\pm}=\pm0.1$ eV/ps, $H_{0}=156.6$ \foreignlanguage{american}{$\textrm{\AA}$},
$W_{0}=9.7$ \foreignlanguage{american}{$\textrm{\AA}$}, $h_{L}=27.9$
\foreignlanguage{american}{$\textrm{\AA}$}, $w_{L}=36.4$ \foreignlanguage{american}{$\textrm{\AA}$},
$h_{R}=54.6$ \foreignlanguage{american}{$\textrm{\AA}$}, $w_{R}=4.0$
\foreignlanguage{american}{$\textrm{\AA}$}. The axis of the beam
is in the x-axis while the axes of the two arms are in the y-axis.
As shown in Fig. 2(a), there is no obvious difference between the
two temperature profiles on the beam by imposing the heat flux $J_{+}$
and $J_{-}$ respectively. Meanwhile, as shown in Fig. 2(b), distinctively
different temperature drops between the two arms are observed by imposing
the heat flux $J_{+}$ and $J_{-}$ respectively. The temperature
drop is much smaller by imposing the heat flux $J_{+}$, therefore
it indicates that the preferential direction is from the left arm
(the wide arm) to the right arm (the narrow arm). The result also
implies that the thermal rectification effect is caused by the asymmetric
arms rather than the beam. 

In order to illustrate the dependence of rectification ratio upon
the heat flux, different heat fluxes are applied. In Fig. 3(a) we
show the heat flux $J$$_{\pm}$ versus the temperature difference
$\bigtriangleup T_{\pm}/T$ between the two arms. It shows that the
decreasing of the temperature difference under $J_{-}$ is more steeply
than the corresponding increasing of the temperature difference under
$J_{+}$. The result indicates again that the U-shaped graphene flake
behaves like good thermal conductor under $J_{+}$ and poor thermal
conductor under $J_{-}$. The heat flux runs preferentially from the
wide arm to the narrow arm. It is similar to the rectification effect
observed in asymmetric graphene nanoribbons where the preferred direction
is along the direction of the decreasing width\citet{11.bb1,12.bb2}.
In Fig. 3(b) we show the quantitative dependence of thermal rectification
ratio upon the heat flux. With $J_{\pm}=\pm0.1$ eV/ps the rectification
ratio is 91\%, while with $J_{\pm}=\pm1.0$ eV/ps the rectification
ratio increases to 104\%. It indicates that although the increasing
of heat flux results in the increasing of rectification ratio, but
the rectification ratio is not very sensitive to the heat flux. The
result demonstrates that the U-shaped graphene flakes have an obvious
advantage in real application. A large rectification effect could
be expected even a small amount of heat flux is generated in the nanoelectronic
devices. 

In order to investigate the influence of the beam upon thermal rectification,
different lengths and widths of the beam are studied. The two arms
are ketp invariant and $J_{\pm}=\pm0.6$ eV/ps are implemented. In
Fig. 4(a) we show that although the rectification effect is enhanced
by decreasing the length of the beam, but it is not very sensitive
to the length of the beam. Even the length of the beam is much longer
than the average length of the two arms ($H_{0}/(\frac{(h_{L}+h_{R})}{2})=4.83$),
the rectification ratio only decreases to 62\%. On the other hand,
in Fig. 4(b) we show that the rectification effect is greatly weakened
by decreasing the width of beam. When the width of beam is small enough
($W_{0}/(\frac{(W_{L}+W_{R})}{2})=0.36$), the rectification ratio
decreases to less than 31\%.%
\begin{figure}
\includegraphics[scale=0.2]{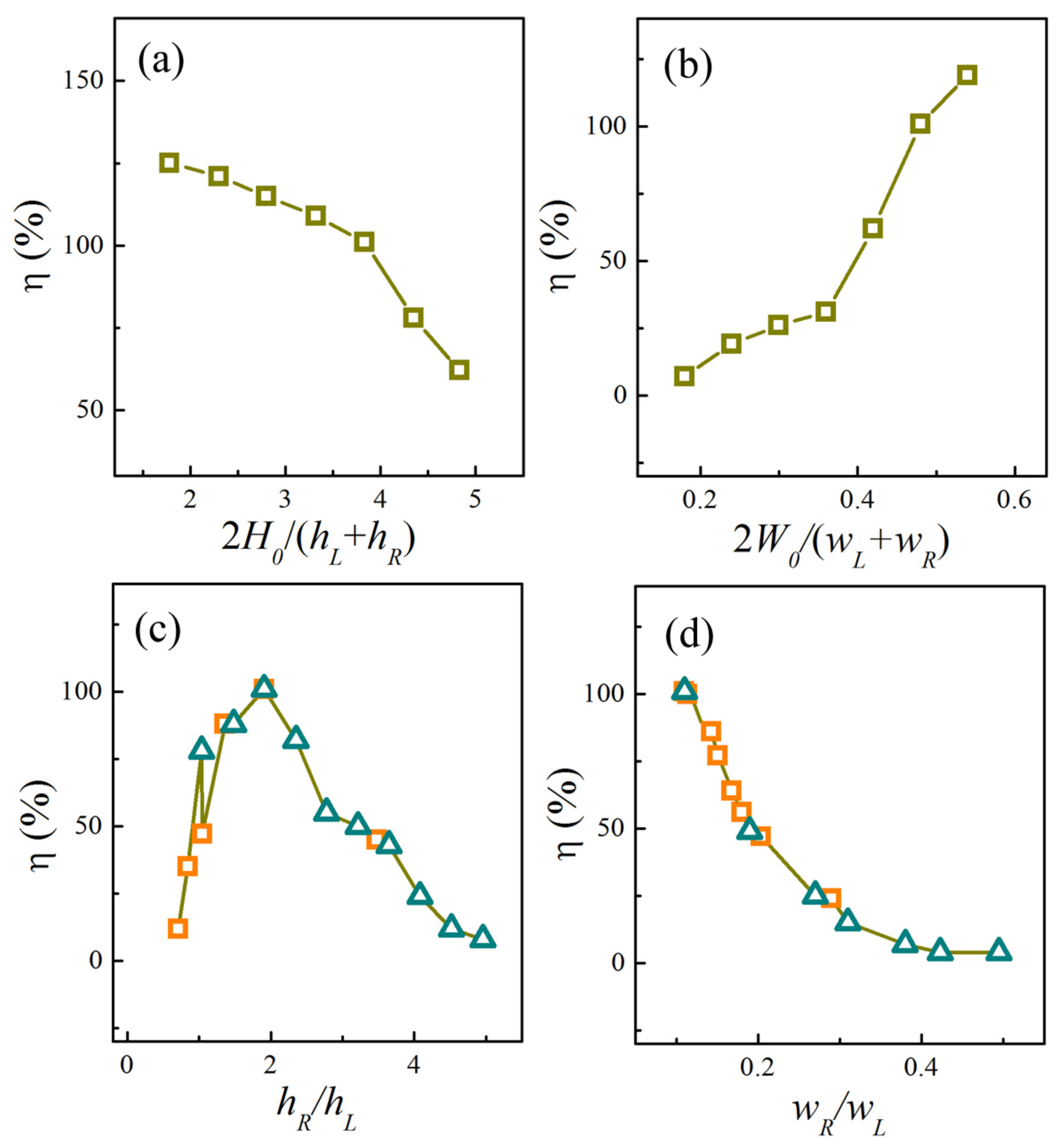}

\caption{(Color online) (a) Rectification ratio $\eta$ vs length of the beam.
$(h_{L}+h_{R})/2=40.9$ \foreignlanguage{american}{$\textrm{\AA}$}
and $H{}_{0}$ varies from 72 \foreignlanguage{american}{$\textrm{\AA}$}
to 197 \foreignlanguage{american}{$\textrm{\AA}$}.(b) Rectification
ratio $\eta$ vs width of the beam.$(w_{L}+w_{R})/2=20.2$ \foreignlanguage{american}{$\textrm{\AA}$}
and $W{}_{0}$ varies from 3.6 \foreignlanguage{american}{$\textrm{\AA}$}
to 10.9 \foreignlanguage{american}{$\textrm{\AA}$}. (c) Rectification
ratio $\eta$ vs length difference ratio between the two arms $h_{R}/h_{L}$.
(d) Rectification ratio $\eta$ vs width difference ratio between
the two arms $w_{R}/w_{L}$. {[}Triangles in (c) and (d) are obtained
by varying the length and width of right arm. Squares in (c) and (d)
are obtained by varying the the length and width of left arm.{]}}

\end{figure}

In order to investigate the effect of structural asymmetry on thermal
rectification, different length and width difference ratio of the
two arms are studied. In Fig. 4(c) we show that the rectification
ratio reaches the maximum value around the length difference ratio
$h_{R}/h_{L}=1.91$. It states that although the length asymmetry
increases with $h_{R}$ or $h_{L}$, the rectification effect would
be weakened by deviating from the proper length difference ratio.
Meanwhile, in Fig. 4(d) we show that the rectification ratio increases
monotonously with the width difference (since $w_{L}>w_{R}$, thus
the width difference is greater when $w_{R}/w_{L}$ is smaller). It
indicates that in real application a large rectification ratio would
be expected by introducing a suitable length asymmetry and a large
width asymmetry between the two arms. 

In summary, we design a thermal rectifier by U-shaped graphene flakes
by introducing asymmetric arms. A strong thermal rectification effect
is observed and the preferred direction of the heat flux is from the
wide arm to the narrow arm. The rectification ratio is not very sensitive
to heatflux which might be important for nanoelectronic devices where
only small amount of heat flux is generated. In addition, we state
that the proper design of the beam and the structural asymmetry between
the two arms are necessary for the rectification effect. Our results
may be useful for engineering graphene based nanoelectronic devices.
\begin{acknowledgments}
We thank Jiao Wang and Yong Zhang for helpful discussion and preparing
of the manuscript. This work was supported by National Natural Science
Foundation of China(\#10775115 and \#10925525).\end{acknowledgments}

\end{document}